\documentclass{article}

\usepackage{arxiv}
\usepackage{textcomp}
\usepackage{xcolor}
\usepackage[utf8]{inputenc}
\usepackage{graphicx}
\usepackage{algorithmic}
\usepackage[linesnumbered,ruled,vlined]{algorithm2e}
\usepackage{amsmath,amssymb,amsfonts}
\usepackage{textcomp}
\usepackage{natbib}
\usepackage[utf8]{inputenc}
\usepackage[T1]{fontenc}
\usepackage{nomencl}
\usepackage{color,soul}
\makenomenclature
\usepackage{xcolor, soul}
\sethlcolor{green}

\title{Internet of Predictable Things (IoPT) Framework to Increase Cyber-Physical System Resiliency}

\author{
 Umit Cali \\
  Department of Electric Power Engineering\\
  Norwegian University of Science and Technology\\ 
  Trondheim, Norway \\
  \texttt{umit.cali@ntnu.no} \\
   \And
 Murat Kuzlu\\
  Department of Engineering Technology,\\
  Old Dominion University, \\
  Norfolk, USA \\
  \texttt{mkuzlu@odu.edu} \\
  \And
 Vinayak~Sharma \\
  University of North Carolina at Charlotte,\\
  Charlotte, NC, USA \\
  \texttt{vsharm12@uncc.edu} \\
  \And
  Manisa~Pipattanasomporn \\
  Smart Grid Research Unit, Chulalongkorn University,\\
  Bangkok, Thailand \\
  \texttt{manisa.pip@chula.ac.th} \\
  \And
  Ferhat Ozgur Catak \\
  Simula Research Laboratory,\\
  Fornebu, Norway \\
  \texttt{ozgur@simula.no} \\
}

\begin{document}
    \maketitle
    
%\corresp{Corresponding author: umit Cali (e-mail: umit.cali@ntnu.no).}

\begin{abstract}
During the last two decades, distributed energy
systems, especially renewable energy sources (RES), have become
more economically viable with increasing market share and penetration levels on power systems. In addition to decarbonization
and decentralization of energy systems, digitalization has also
become very important. The use of artificial intelligence (AI),
advanced optimization algorithms, Industrial Internet of Things
(IIoT), and other digitalization frameworks makes modern power
system assets more intelligent, while vulnerable to cybersecurity
risks. This paper proposes the concept of the Internet of Predictable Things (IoPT) that incorporates advanced data analytics and machine learning methods to increase the resiliency of
cyber-physical systems against cybersecurity risks. The proposed
concept is demonstrated using a cyber-physical system testbed
under a variety of cyber attack scenarios as a proof of concept
(PoC).
\end{abstract}

%\begin{keywords}
%Internet of Predictable Things (IoPT), energy forecasting, machine learning, edge analytics\\
%\end{keywords}

%\titlepgskip=-15pt

\section{ Introduction }
The liberalization process of the energy sector and global Organization of the Petroleum Exporting Countries (OPEC) crisis in the 1970s are two major drivers of the decentralization and decarbonization energy generation systems. Distributed energy systems, especially renewable energy sources (RES), have become more economically viable, and their market share has significantly increased in the last two decades.  Wind and solar energy plants are the most prominent RES, which generates a fluctuating and weather dependent power output. Power systems are operated according to certain national and international norms where the voltage and frequency parameters should not exceed certain operational boundaries. Power networks are also designed to carry specific maximum power capacities. The power output characteristics of RES increase the vulnerability and uncertainty levels of power systems, which makes it challenging for the power systems operators to integrate higher amounts of RES into their control zones. Energy forecasting is one of the most promising methods which increases the operational capabilities of RES. 
Wind and solar forecasting algorithms have been used for two decades by various energy market players, such as utilities, RES plant operators, and power traders. Transmission and distribution system operators use energy forecasting algorithms to schedule their daily energy generation profiles, thus minimizing last-minute balancing power needs. Wind and solar plant operators use similar forecasting tools to optimize their power generation assets for efficient maintenance and power trading operations. Future energy markets will accommodate new players such as virtual power plant operators, regional power aggregators, and autonomous prosumer agents. Customer or prosumer edge is expected to host a variety of innovative, intelligent energy use cases and business models in the future. Recent advances in artificial intelligence, the Internet of Things (IoT), communication, as well as distributed energy systems enable researchers to develop new cyber-physical system technologies, hence making energy systems more intelligent and secure towards the industry 5.0 era.
Renewable energy generation, such as solar PV and wind energy, is highly weather dependent. This makes the generation from such energy sources uncertain and variable in nature. Techniques to forecast solar PV and wind power have been extensively discussed in the literature. One of the most basic forecasting techniques is the persistence forecasting, which has been widely used as a benchmarking technique to compare with
\citep{fernandez2012short, monteiro2013short, cali2011grid, biermann2005entwicklung, cali2008artificial}. However, since solar PV power generation is highly weather dependent, it is common to use numerical weather prediction (NWP) data to forecast solar PV power. Instead of forecasting solar PV power directly, some papers propose to forecast solar irradiance and subsequently calculate solar PV power \citep{hammer1999short, mellit201024}. Models that are statistical, such as auto-regressive (AR) models, auto-regressive moving average (ARMA) models, as well as exogenous variants, are known to provide reasonable forecast \citep{bacher2009online,bessa2015probabilistic}. Autoregressive integrated moving average (ARIMA) and seasonal ARIMA models have also been explored \citep{reikard2009predicting,pedro2012assessment}. Machine learning models, such as Support Vector Regression (SVR) \citep{alfadda2017hour}, Artificial Neural Networks (ANN) \citep{chen2011online,ding2011ann}, Random Forest (RF) \citep{zamo2014benchmark}, Non-linear Autoregressive Artificial Neural Network \citep{tao2010forecasting,sharma2018numerical} discussed in the literature \citep{antonanzas2016review} to classify days into various day types based on weather, such as foggy day, rainy day, sunny day, etc. Different models are created for each day type, and later all the models are combined in order to get the final solar PV power forecast. Models without using any NWP information to predict short-term solar PV power have also been explored, when obtaining the NWP is not possible due to emergency situations like communication failure \citep{sharma2018deterministic,ordiano2017photovoltaic}.

Electricity demand forecasting is one of the most critical inputs in planning and decision making in the energy sector.  The extensive amount of research can be found in the literature on short-term, medium-term, and long-term load forecasting \citep{jia2001flexible,suganthi2012energy}. Various techniques have been used to forecast electricity demand, such as linear regression \citep{song2005short}, neural networks \citep{hippert2001neural}, random forest \citep{lahouar2015day}. Deep learning ensemble methodologies have also been explored in the literature \citep{qiu2014ensemble}. Due to the high amount of renewable penetration at the utility level as well as the behind-the-meter level, the traditional load curve is changing. For this reason, net-load forecasting becomes an important step in planning for energy demand. The net load is referred to as the demand for electricity after subtracting renewable generation \citep{kaur2016net}.

Cybersecurity is an emerging challenge for power systems since the impact of cyber attack incidents on the power system has suddenly been seen on the grid operation, such as socioeconomic impacts, market impacts, equipment damage, and large-scale blackouts \citep{carreras2002critical,qi2013blackout}. Cybersecurity of energy forecasting systems, especially for RES, is also very crucial to prevent, detect, and respond to an attack, such as manipulated RES data and operation. RES outputs strongly depend on weather conditions, and it is difficult to accurately estimate and predict RES operations for early detection of anomalies. Therefore, it is better to predict the RES operation and anomalies by data-driven approaches and advanced data analytics to detect potential cyber attacks.  

\textcolor{black}{This paper presents a new concept, which uses a hybrid CPS framework consisting of IoT, artificial intelligence, edge analytics, and cybersecurity components together, as well as introduces new terminology called the Internet of Predictable Things (IoPT). Several use cases are also discussed, including forecasting systems under a variety of cyberattack scenarios. One objective of the proposed concept is to create an adaptive context framework to improve the grid operations and forecasting accuracy while eliminating possible cyber-attacks.  This is intending to investigate them and measure the performance of the developed algorithms. The results show that the proposed concept can reduce the surface of cyberattacks, obviously. One of the most critical vulnerabilities in an IoT environment is an adversarial machine learning attack. Attackers target machine learning models in IoT environments as a new attack surface for the poisoning or evading them. Most of the adversarial machine learning attacks rely on adding noise to the input instances in the training or testing phase. Although there are some noise adding based sophisticated adversarial machine learning attacks, such as Fast Gradient Sign Method (FGSM) \citep{goodfellow2014explaining}, Basic Iterative Method (BIM) \citep{kurakin2016adversarial}, DeepFool \citep{moosavi2016deepfool}, we aim to increase the robustness against the gaussian noise-based attacks which are compatible with the current literature in this work.}

\section{Framework For Internet of Predictable Things}
In a typical scenario, IoT devices are used as sensors and gateway devices in the system. Data are constantly collected from "Things", and passed on to a central hub via a gateway. The data analysis and preprocessing are performed in the central hub using the state of the art applications and software to create an intelligent system. 

This paper proposes the framework that combines IoT sensors, IoT gateways, and analysis in one device, called Internet of Predictable Things (IoPT), as shown in Fig. 1. The physical layer includes physical components of CPS, such as sensors, hardware components of IIoT and any physical elements of power systems.  The communication layer is responsible for hosting various communication protocols and technologies. The security layer accommodates the cybersecurity algorithms. The application layer enables data analytics functionalities on the edge, in other words, on IoT devices. Data or edge analytics elements can accommodate machine learning or artificial intelligence scripts depending on a specific application.  IoT gateways enable data fusion and link various clusters of IoT or IoPTs together. Optionally the cloud network can also be used in the ecosystem.     

%\Figure[!t](topskip=0pt, botskip=0pt, midskip=0pt)[width = 85mm]{Fig_1.png}
%{Components of the proposed Internet of Predictable Things (IoPT) concept.\label{Fig:fig1}}

\begin{figure}
	\includegraphics[width=1\linewidth]{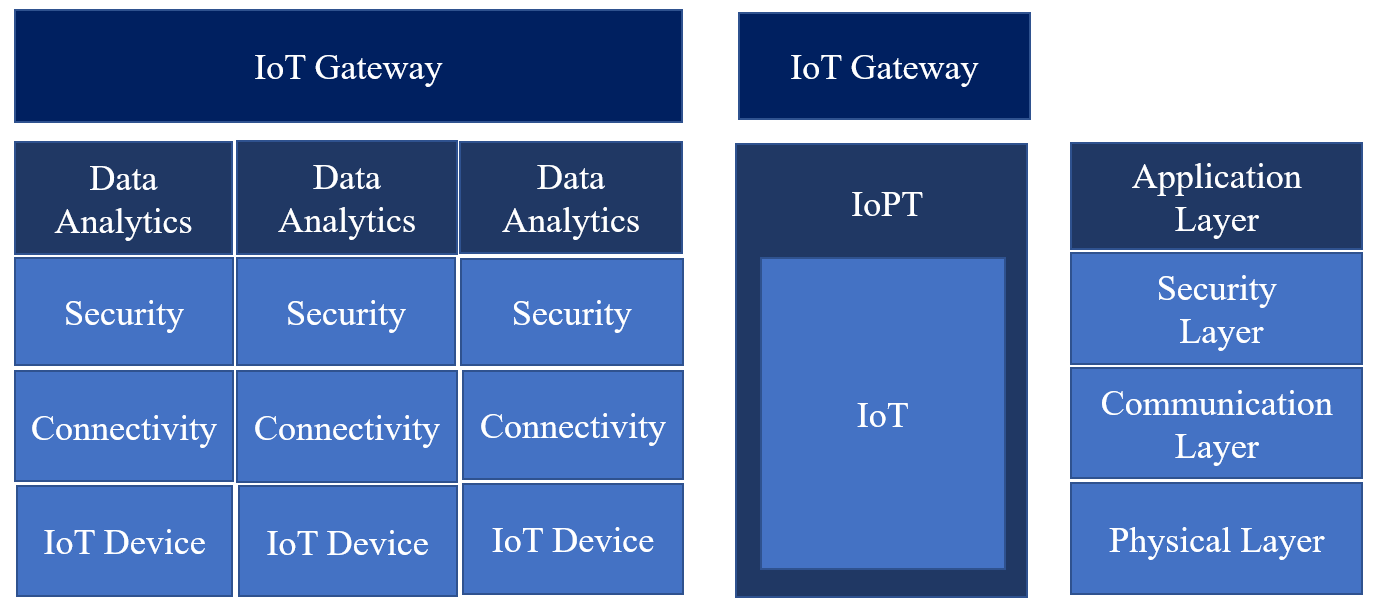}
	\caption{Components of the proposed Internet of Predictable Things (IoPT) concept.}
	\label{Fig:fig1}
\end{figure}

The main contributions of this work are listed below: 
\begin{enumerate}
\item  A new hybrid CPS technology approach is proposed, which combines advanced data science, IoT, edge analytics, and intelligent industrial systems together, called IoPT. 
\item  Technology mapping of the new concept is explained. 
\item  IoPT-functional proof of concept using machine learning-based net-load forecasting algorithms is successfully deployed and tested on an IoT platform. 
\item  \textcolor{black}{Demonstration of improved cyber-physical security resiliency using the IoPT concept against noise based adversarial machine learning attacks as an initial case scenario.} \end{enumerate}

\begin{figure*}[!ht]
%\begin{figure}[!ht]
  \centering   
  \includegraphics[width=150mm,height=100mm,keepaspectratio]{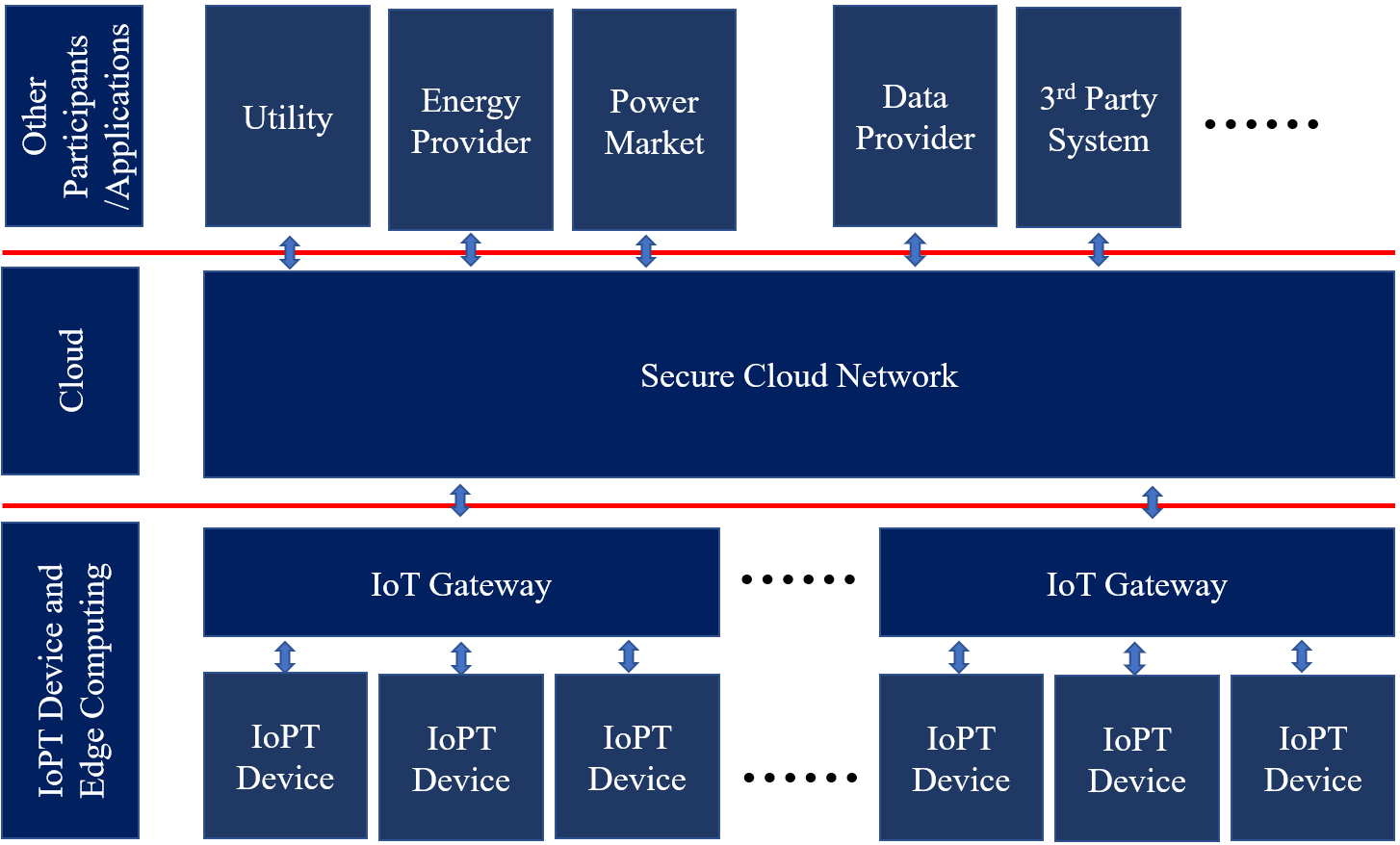}
  \caption{The cloud-based IoPT architecture consisting of a cloud network, a set of machine learning algorithms and IoT devices for the energy use cases}
\end{figure*}

\section{Description of the Proof of Concept IoPT}
The proposed secure architecture IoPT framework is a novel edge analytic and optionally a cloud-based architecture, which embeds a set of artificial intelligence-driven algorithms on an IIoT edge. The IoPT framework can be applied to smart cities, energy, health, and digital logistics fields. Fig. 2. depicts the proposed cloud-based IoPT architecture for the energy sector use cases. Actors of the energy landscape, such as Utility, Energy Provider, Power Market participants and others connected each other via the secure cloud network. IoPT devices and other edge computing elements are connected to the secure cloud network via IoT gateways. 

\textcolor{black}{It is assumed that the secure communications between actors of the energy landscape and IoPT devices will be carried out through encrypted links using SSL/TLS protocols to avoid possible vulnerabilities, especially for the data disclosure problems in real world implementations. In this study, the adversarial machine learning attacks and cyber-attack are used interchangeably.}

This paper demonstrates the proposed approach in the energy field using the use case of net load forecasting that incorporates energy, climate, and cyberattack-resilient components. It consists of a cloud network, a set of machine learning algorithms, i.e., edge learning, and IoT devices, i.e., sensors and gateway devices. The proposed concept aims at collecting data from IoT devices and third-party data providers, such as meteorology service providers (numerical weather prediction service providers) through IoT gateways. It is assumed that the collected data include time-series data (building-level and utility-level), and public data. Building-level data include energy usage, renewable energy generation, and storage (optional), and utility-level data include demand response (DR) event, pricing, and power data. These data are gathered from utilities or public and commercial data sources. Weather data, satellite images and social activities are collected through publicly available data sources. Then the collected data are pre-processed and utilized by net load forecasting algorithms. The net-load forecasting algorithms gather generation data from DER (e.g., solar PV power generation), electricity consumption from loads, and calculate the residual. Cybersecurity scenarios include detection and mitigation of anomalies in forecasting models’ input data, which can seriously affect forecasting results.  The framework features various AI/ML techniques, data sources, and data generation models. 

Fig 3. demonstrates more details about the developed PoC, where some of the main functionalities of the IoPT framework are tested. Energy forecasting tools and algorithms can be executed in various locations of power systems and the market value chain, such as transmissions system operators (TSOs), distribution system operators (DSOs), control rooms of the power generation asset operators, power traders and retail energy providers. Optionally energy forecasting service providers can also run the energy forecasting tools and send the forecasted energy time-series data to their clients. In the presented use case, an IoPT-based forecaster is deployed on a grid edge, where a prosumer is located. Stakeholders marked with blue star represent the locations where classical energy forecasting algorithm or tools can be executed. IoPT based energy forecaster is marked with the red star. Experiments are performed using Raspberry Pi, where the real ML-based net load forecasting algorithm is executed. 

\subsection{Edge Learning: IoPT as an Addition to Edge Computing }

\begin{figure}
  \centering    
  \includegraphics[width=90mm,height=100mm,keepaspectratio]{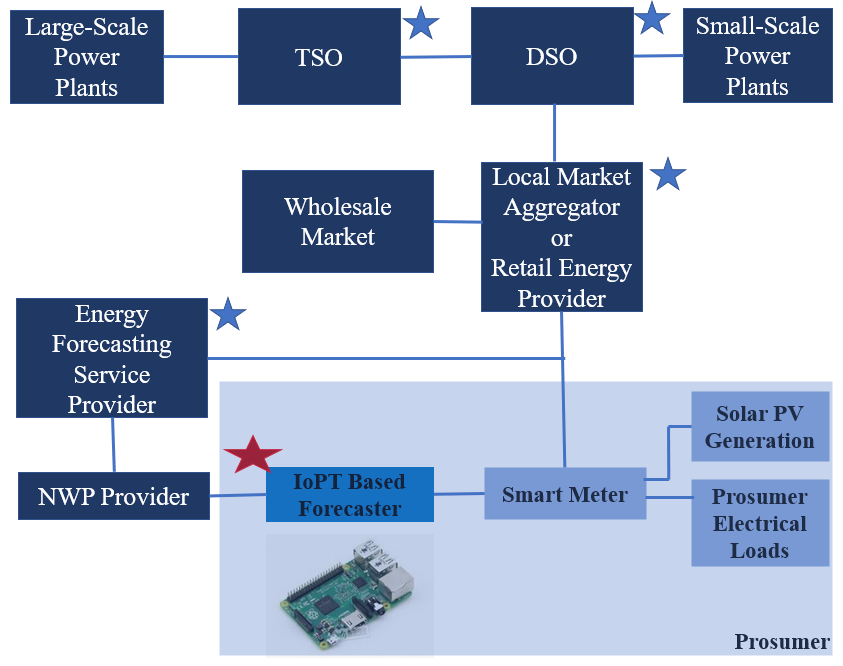}
  \caption{Overview of experimental PoC of IoPT based energy forecasting}
\end{figure}

A set of IoT devices comprise a network of sensors and embedded devices connected to the Internet. Data are collected by embedded devices, called "Things." This data is transferred on to the Internet via an IoT gateway. In the current scenario, IoT devices that can be seen in energy systems collect real-time electricity and environmental data, such as power meters, temperature sensors, etc. Data collected from these sensors are transferred to the cloud via an IoT gateway where the data are further analyzed \citep{jaradat2015internet}. 

As an extension of the IoT-based power system model, computation capability can be added to the edge of the network as compared to the entire computation taking place in the cloud. In such cases, edge devices are tasked to perform some computations, such as detecting anomaly, pre- processing the data and caching. This reduces the computational load on the cloud services~\citep{shi2016promise}.  
 
The proposed IoTP concept adds another dimension to the concept of edge computing, which we call "Edge Learning." That is, in addition to collecting and processing data, edge devices can also take care of the entire computation load. With this concept, traditional IoT devices (such as a sensor that can only transmit its reading, i.e., one-way data stream) can now combine the capability of a sensor, a gateway, and perform the machine learning and data analysis tasks locally, with a bi-direction data stream. This proposed methodology is expected lead to the following advantages:

\begin{enumerate}
\item  Less computation load on the cloud.
\item  Quick response to changes without any latency.
\item  Less vulnerability to cyber attacks.
\item  Reduce the dependence on central forecasting systems.
\end{enumerate}

\subsection{Net Load Forecasting}
The proposed net load forecasting model utilizes solar PV generation and electricity consumption data as input variables. Since the objective of this article is not to propose a new method for net load forecasting, a functional net load forecasting method based on a neural network is utilized for the proposed PoC. As shown in Fig. 4, a hybrid net load energy forecasting model is developed to demonstrate the applicability of IoPT. It has also been tested to evaluate the impact of cyber attacks on the proposed concept.   

\subsection{Load Forecasting: A Neural Network Based Approach}
Neural network models try to learn relationships between inputs and outputs by mimicking the human brain. Neural networks have been commonly used in forecasting electricity load due to their ability to learn non-linear relationships~\citep{park1991electric}. A neural networks model is made up of neurons arranged in different logical layers, namely the input layer, hidden layer, and output layer, connected to each other via weighed connections~\citep{lippmann1987introduction}. The neural network model learns by updating and adjusting the weights of the interconnections. The model adjusts its weights by comparing the target value with the computed value in each run. This optimization of weights to achieve minimum error is done by a method known as back propagation~\citep{goh1995back}. Mathematically a neural network model can be given as:
\begin{equation}\mathrm{Y}_{\mathrm{j}}=\mathrm{f} \sum \mathrm{iw}_{\mathrm{ij}} \mathrm{X}_{\mathrm{ij}}
\end{equation}
Where, $Y_j$ is the output of the neuron $j$, $X_ij$ is the input from neuron $i$ to $j$ and $W_ij$ is the weight between the two neurons. The function $f$ is known as the transfer function. Sigmoid is the most commonly used transfer function, given by:
\begin{equation}
y=\frac{1}{1+e^{-x}}
\end{equation}
The back-propagation algorithm tries to minimize the sum of errors between the target ($O_j$) and the output from the neuron, given by:
\begin{equation}
E=\frac{1}{2}\left(O_{j}-Y_{j}\right)^{2}
\end{equation}
Once the output is calculated, the gradient descent algorithm tries to adjust its weights according to the steepest direction~\citep{hsu2003regional}. This process takes a few iterations (k), and at the end, the updated weights can be given as:
\begin{equation}
\Delta \omega_{\mathrm{ji}}(\mathrm{k})=-\eta(\mathrm{k}) \partial \mathrm{E}(\mathrm{k}) \partial \omega_{\mathrm{ji}}(\mathrm{k})
\end{equation}
Where  $\eta(\mathrm{k})$ is the learning rate that is set at the beginning of the training process~\citep{huang2002new}.
%\begin{equation}
%\hat{O_{m}}=\frac{1}{1+\exp \left(-\sum_{j=1}^{h} W_{j m} \hat{HL_{j}}\right)}
%\end{equation}
%where $\hat{O_{m}}$ is the estimated output. $\hat{HL_{j}}$ is the output of the previous hidden %layer connected to the output layer by the weighted connection $W_{j m}$.
%$\hat{HL_{j}}$ can be given as:
%\begin{equation}
%\hat{HL_{j}}=\frac{1}{1+\exp \left(-\sum_{i=1}^{n} W_{i j} X_{i}\right)}
%\end{equation}
%where $X_{i}$ is the input~\citep{khotanzad1998annstlf}.

In this study, ANN was used as the learning model for forecasting electrical load. The dataset was divided into training (70\%) and testing (30\%) sets. The training set was fed into the ANN model. The model learned the relationship between the inputs and the target variable to obtain a trained model. The temperature values from the testing data were then fed into the trained model, and the prediction was compared with the corresponding load data from the testing set to evaluate the model accuracy. This study used an ANN model with one hidden layer and two hundred hidden neurons. These values were decided after conducting experiments with different values of hidden layers and hidden neurons.

\subsection{Solar PV Forecasting: A Gradient Boosting Based Approach}
Gradient boosting machines are a powerful machine learning algorithm that combines gradient descent and boosting to perform regression as well as classification tasks. It was introduced in \citep{friedman2001greedy}. Gradient boosting combines base learners in the steepest direction via numerical optimization. The resultant of each base learner is an estimation of the output given by $\hat{f}_{i}(x)$. 
\begin{equation}
\hat{Y}(x)=\sum_{i=1}^{I} \hat{f}_{i}(x)
\end{equation}
At each iteration, the gradient boosting algorithm calculates the steepest direction or the direction of minimum error, given by the negative gradient of $D(\mathbf{f})$, defined as:
\begin{equation}
D(\mathbf{f})=\sum_{i=1}^{N} \Psi\left(y_{i}, f\left(\mathbf{x}_{i}\right)\right)
\end{equation}
The function is updated as:
\begin{equation}
\hat{\mathbf{f}} = \hat{\mathbf{f}}-\delta \nabla D(\mathbf{f})
\end{equation}
Where, $\delta$ is the step-size and  D({f} is the steepest direction~\citep{ridgeway2007generalized,persson2017multi}.
$i$ represents the number of base estimators used in the model~\citep{bessa2014spatial}.
\textcolor{black}{PV data used in the paper comes from the Global Energy Forecasting Competition 2014. The data includes weather variables like irradiance, temperature, wind speed, and other relevant variables which were used as inputs to the forecasting model along with additional variables such as day of the week, the month of the year, the hour of the day, etc. These added variables were encoded using one-hot-encoding those are resulting in more columns. In total, there were 350 parameters in the model that is presented in this study.} For this work, the number of estimators of 350 was selected. Similar to the load forecasting, the dataset was divided into training (70\%)  and testing (30\%) sets. The inputs used in the model consist of historical PV power data and corresponding numerical weather prediction variables, namely, temperature.

\section{Scenario and Data Description}

\subsection{Load and Solar PV Data Collection and Prepossessing}
The load data used for the experiments are an open-source dataset from the Global Energy Forecasting Competition held in 2012. The dataset consists of four and half years of historical hourly load data from 1/1/2004 to 6/30/2008 for 20 zones plus an aggregated zone marked as zone 21. The corresponding temperature data from 11 different weather stations are available in the dataset \citep{hong2014global}. 

In this paper, the electricity load data from the aggregated zone 21 was used. For the temperature data, a virtual weather station was created by taking a simple average of the 11 temperature values provided. Calendar variables such as the month of the year, day of the week, holiday and day of the year are used as additional inputs to enhance the accuracy of the model. Data from the solar forecasting track of the Global Energy Forecasting Competition 2014 \citep{hong2016probabilistic} were used in this paper. The open-source data-set contains three years of hourly solar generation and corresponding weather data from 4/01/2012 to 04/06/2014. Fig. 4 shows proposed hybrid IoPT based energy forecasting model.

\begin{figure}[!ht]
  \centering  
  \includegraphics[width=90mm,height=100mm,keepaspectratio]{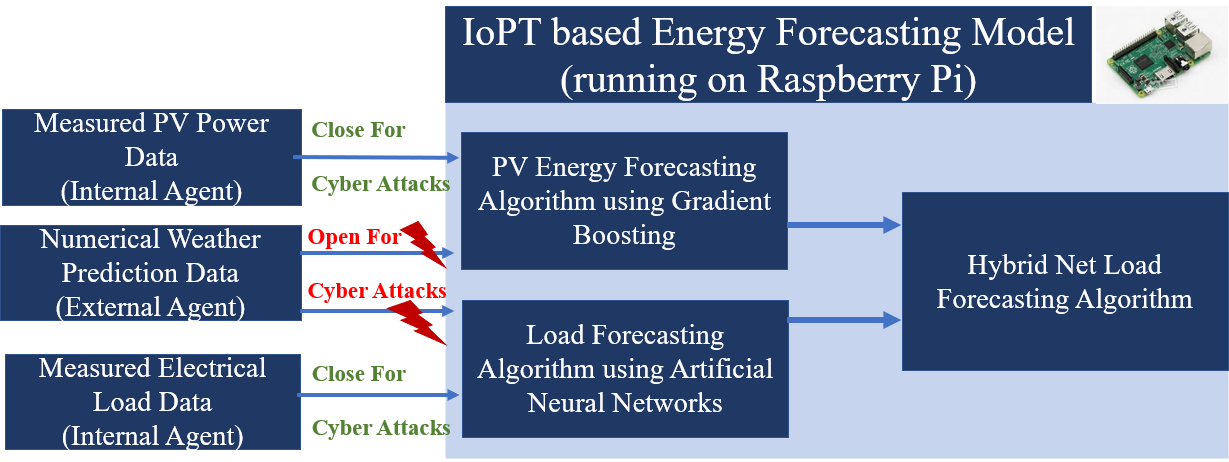}
  \caption{Hybrid IoPT based energy forecasting model}
\end{figure}

\subsection{Performance metrics}
Root Mean Squared Error (RMSE) and Mean Absolute Percentage Error (MAPE) were the two metrics used for evaluating the performance of the developed forecasting models. RMSE and MAPE are the two most popular error metrics used in the literature for forecasting evaluation. RMSE and MAPE are defined as:

\begin{equation}\label{MAE}
MAPE= \frac{1}{N}\frac{\sum\limits_{n=1}^{N} |y[(n)-\hat{y}(n)|}{y(n)}
\end{equation}

\begin{equation}\label{RMSE}
RMSE= \sqrt{\frac{\sum\limits_{n=1}^{N}(y(n)-\hat{y}(n))^2}{N}}
\end{equation}

Where $y(n)$ is the actual value at time-step $n$, $\hat{y}(n)$ is the forecast value at $n$ and $N$ is the number of time-steps.

\subsection{Experiments with Various Cyber Attack Scenarios}
In this paper, forecasting systems under various cyber attacks were investigated, and the corresponding net load forecasting accuracy was measured. Cyber attacks were simulated in different ways to create real-world scenarios, where each part of the system could be under attack separately or all at once, as summarized in Table I.

%Table 
\begin{table}
\centering
\caption{Use cases with details}
\begin{tabular}{lllll}
\textbf{Experiment } & \textbf{Experiment} & \textbf{Noise} & \textbf{Noise} & \textbf{$\sigma$; $\mu$; $\%$} \\ 
\textbf{Number}      & \textbf{Domain}     & \textbf{on NWP}& \textbf{on Load} & \textbf{}\\  
\textbf{Base Use Case}  &Central/IoPT &N/A &N/A  &N/A\\
\textbf{1a} &Central &N/A &Testing+  &50;10;10\\
\textbf{}   &        &    &Training  &\\
\textbf{1b} &Central &N/A &Training  &50;10;10\\
\textbf{2a} &Central/IoPT &Testing+ &N/A  &50;10;10\\
\textbf{}   &             &Training &     &\\
\textbf{2b} &Central/IoPT &Testing &N/A  &50;10;10\\
\textbf{3a} &Central &Testing+ &Testing+  &50;10;10\\
\textbf{}   &        &Training &Training  &\\
\textbf{3b} &Central &Testing+ &Training  &50;10;10\\
\textbf{}   &        &Training &          &\\
\end{tabular}
\end{table}

Firstly, the system in a normal operating condition, i.e., no cyber-attack, was simulated. This experiment is called experiment 0 or the base case. It acts as a benchmark for further experiments. \textcolor{black}{To mimic a cyberattack (an adversarial machine learning attack),} noises generated based on a methodology, as described in~\citep{luo2018benchmarking}, were injected into the dataset. In particular, cyber attacks were simulated by randomly selecting 10\% of the data to inject distributed noises with a mean ($\mu$) of 10 and a standard deviation ($\sigma$) of 50. This approach mimicks the data integrity type of cyber attacks.

In experiment 1a, noises were added to the entire (train + test) load dataset. In experiment 1b, normally distributed noises were added just to the training set of the load data. \textcolor{black}{This is a typical data poisoning attack to fool the machine learning models.} This delivers a more realistic scenario where the historical load might get affected by an attack on the data. The experiments 1a and 1b were executed on the central domain, representing a dedicated server or a cloud computing environment. In the next set of experiments, noises were added to the NWP data. In experiment 2a, noises were added to the entire (train + test) NWP dataset. This mimics the situation wherein the NWP data source might get affected by an attack. In experiment    2b, the only testing set of the NWP data was affected by noises. \textcolor{black}{This is a typical model evading attack to fool the machine learning models.} This mimics a scenario wherein; the historical NWP data is not affected, but the current NWP data is manipulated. IoPT experiments can be executed for experiments 2a and 2b besides the central domain experiments. In experiment 3a, the entire (train + test) dataset of NWP data, as well as the load data, were affected. In experiment 3b, the entire NWP dataset and the training dataset of load data were attacked. All the described experiments mimic different real-world scenarios above where the forecasting system could be affected by cyber-attacks.

\begin{figure}[!ht]
\centering  
\includegraphics[width=90mm,height=100mm,keepaspectratio]{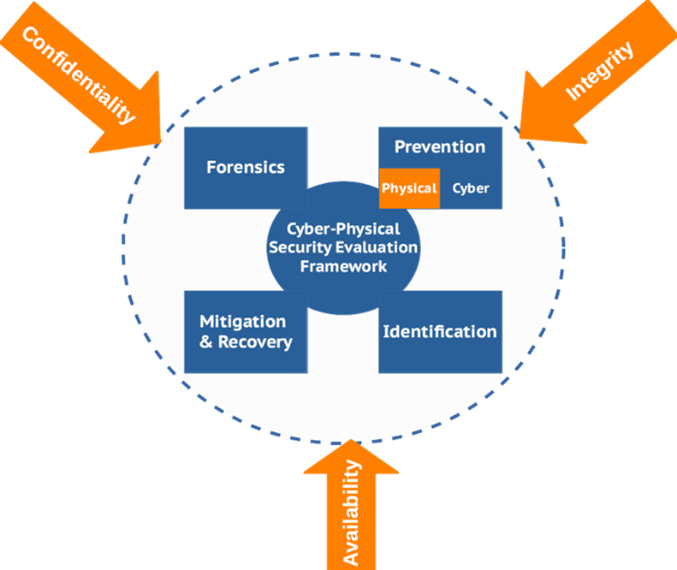}
\caption{Holistic cyber-physical security evaluation framework}
\end{figure}

%\begin{figure*}[!ht]
%  \centering    
%  \includegraphics[width=175mm,height=110mm,kee%paspectratio]{Fig_5.png}
%  \caption{(a) Solar PV power actual/forecasted w/ and w/o Noise and (b) Irradiance and Noise}
%\end{figure*}

\begin{figure*}[!ht]
\centering    
\includegraphics[width=175mm,height=110mm,keepaspectratio]{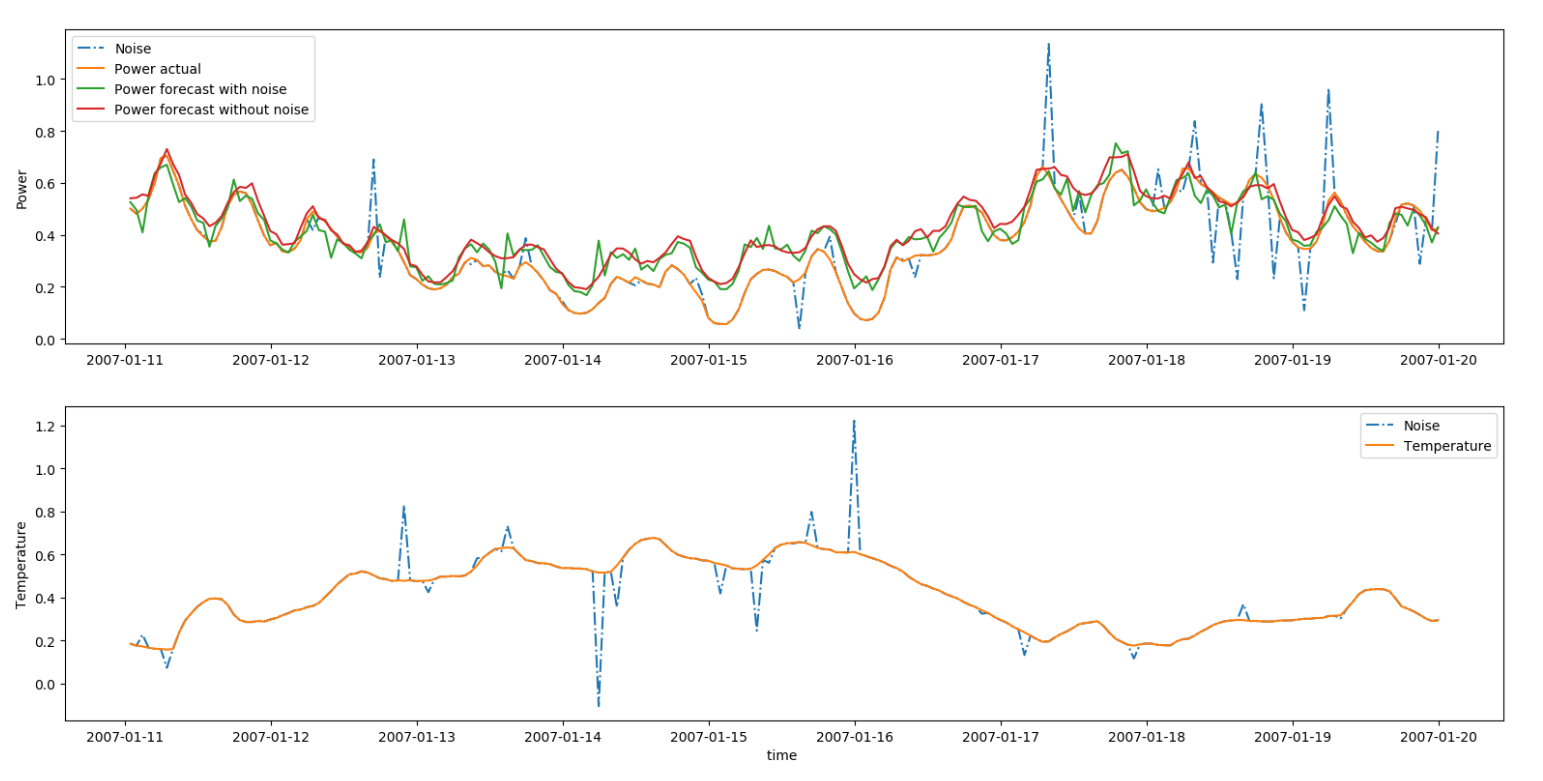}
\caption{(a) Net power actual/forecasted w/ and w/o Noise and (b) Temperature and Noise}
\end{figure*}

\textcolor{black}{
\subsection{Cyber-Physical System Security Evaluation}
The well-known cyber-physical security framework in cybersecurity is confidentiality, integrity, availability (CIA) triad \citep{maalem2020review}.  The CIA approach is a holistic method, which can be implemented to any type of cyber-physical security infrastructure framework, as shown in Fig. 5. Confidentiality guarantees that sensitive data are accessed only by the authorized person and protected from the users who are not allowed to access them. Integrity is responsible for enabling certain features so that data are processed and stored properly, compliant with its initial design parameters and constraints in terms of correctness, reliability, and fidelity. The data can be edited by only authorized persons and persists in its primary state when at rest. Availability ensures that data and computation resources like network, CPU, GPU, memory are available to legitimate users with smooth operation in terms of software and operating system execution on the target hardware. In this study, it is considered that the CIA Triad approach can be deployed to the proposed IoPT concept by covering some segments such as physical security and adversarial robustness. 
Data is an asset that needs proper protection, which can also be clustered and classified in terms of importance levels and value propositions for the corresponding organizations such as individual research institutes or companies. Cyber-physical security measures must be applied to protect data from unintentional or intentional illegal alteration, damage, disclosure, or similar activities. 
The prevention phase proposes, designs, and implements security policies, controls, and processes. Prevention techniques can be classified as physical and cyber-related solutions. In the IoPT approach, the physical security prevention convention is demonstrated. 
The well-known physical access IoT attacks are invasive and non-invasive \citep{rizvi2020identifying}. A non-invasive attack does not need any initial arrangements of the IoT device under attack, and will not physically change the IoT device during the attack. The attacker will be very close to the device to sense electrical characteristics to gather sensitive information. For the invasive attack, the attacker needs to access to the chip surface to manipulate the device physically. In this study, physical prevention methods had been investigated to minimize the risk of cyber-physical attack by locating the IoPT device where some components of needed data-streams are connected to the system from an internal communication domain or network. As demonstrated in Fig. 4, the numerical weather prediction data provider channel is open for cyber attacks since it is categorized as an external data-stream source but the measured PV power and electrical load data can be provided from the internal communication domain where the cyber-attack risk is an approximation to zero. This physical prevention method may reduce the attack surface almost down to 50\% lower levels depending on the individual scenario followed in the real-world application. In classical use cases of energy forecasting systems are more open to cyber-physical attacks than the described scenario above. 
Even though this article's main scope is not increasing the accuracy of the energy forecasting algorithms and diversifying the possible cyber-physical security scenarios, it is intended to demonstrate an applicable IoPT framework as a new concept under edge analytics and show-case a functional proof-of-concept with some limitations. In this study, we analyzed the security threats with two different aspects. Machine learning models are vulnerable to the designed adversarial input instances by adding carefully designed noise in training and testing phases of the functional energy forecasting algorithms \citep{pathak2020security}. This study showcases the implementation of adversarial input attacks scenarios and its consequences in terms of energy forecasting accuracy besides investigating the physical prevention of cyber-attacks strategy to increase the system's resilience.}

\section{Result and Discussion}

This section demonstrates the applicability of the proposed IoPT concept for net load forecasting and cyber resilience. We executed the total of six different scenarios by using the simulation tool developed in Python. The simulation tool consists of the net load forecasting algorithm, a noise adding module, and a cyber attack detection algorithm. 

The accuracy of the load forecasting model without any cyber attacks is 5.34 $\%$ (MAPE) for the base case. Experiments 3a and 3b yield the highest MAPE values of 8.57 $\%$ and 8.43 $\%$, respectively, where an intensive data integrity cyber attack was emulated. The MAPE values of 1a and 1b are 7.7 $\%$ and 5.82 $\%$, respectively. Experiments 2a and 2b demonstrate impacts of the IoPT framework on the load forecasting accuracy, which result in MAPE of 6.42 $\%$ and 5.8 $\%$, respectively. As shown in Fig. 3, IoPT based Energy Forecaster is connected to an NWP provider (external agent) and a smart meter or energy gateway (internal agent). While the NWP provider is an open target for possible cybersecurity attacks, a smart meter, which provides the measured power consumption data can also be vulnerable to external hacking attempts due to the physical location of its connections. When comparing experiments 2a and 3a, it is possible to improve the forecasting accuracy and decrease the impacts of possible cyber attacks by 31.1 $\%$ where the potential surface of the attacks can be reduced when implementing the IoPT framework in the architecture. 

The forecasting accuracy of the solar PV forecasting model without any cyber attacks is 7.9 (RMSE) for the base case. This experiment delivers the best results with the lowest RMSE value. Experiments 3a and 3b yield the highest RMSE values of 10.1 and 9.86, respectively, where intensive data integrity cyber attack was emulated. The RMSE values in experiments 1a and 1b are 9.7 and 8.69, respectively. Experiments 2a and 2b demonstrate the impacts of the IoPT framework on the load forecasting accuracy, which results in RMSE of 8.6 and 8.3, respectively. Similar to the load forecasting modeling, the NWP provider is an open target for possible cyber security attacks, and the smart meter, which provides the measured solar PV generation values, is vulnerable to external cyber attack actions due to the physical location of its connections. When comparing experiments 2a and 3a, it is possible to improve the forecasting accuracy and decrease the impacts of possible cyber attacks by 17.4 $\%$ due to the IoPT framework. Fig. 6 demonstrates the impact of data integrity attacks (injected noises to the dataset) on the net load forecasting output. 

\begin{table}
\centering
\caption{Load and PV power forecasting results}
\begin{tabular}{llllll}
\textbf{Experiment } &  \textbf{Load forecasting
results} & \textbf{PV power forecasting results} \\ 
\textbf{Number}      & \textbf{MAPE \%} & \textbf{RMSE} \\  
\textbf{Base Case}   & 5.34 & 7.90 \\
\textbf{1a} & 7.7 & 9.70  \\
\textbf{1b} & 5.82 & 8.69  \\
\textbf{2a} & 6.42 & 8.60  \\
\textbf{2b} & 5.80 & 8.30  \\
\textbf{3a} & 8.57 & 10.10 \\
\textbf{3b} & 8.43 & 9.86  \\
\end{tabular}
\end{table}

\section{Conclusion}
This paper presents the new concept of the hybrid CPS framework, consisting of IoT, advanced edge data analytics, and cyber security components, as well as introduces a new terminology to the literature, namely Internet of Predictable Things (IoPT). In addition, energy forecasting systems under various cyber attacks were investigated, and the corresponding net load forecasting accuracy was evaluated. Cyber attacks were simulated in different ways to mimic real-world scenarios, where each part of the system could be under attack separately or all at once. The IoPT concept is demonstrated using a cyber-physical system (CPS) testbed under various cyber attack scenarios as a proof of concept. According to the findings of this study, the proposed IoPT-functional proof of concept with machine learning-based net load forecasting algorithms can reduce the cyber attack related consequences even without additional cyber defense software solutions, such as anti-virus. \textcolor{black}{In addition, it has a high potential to improve the quality of grid operations by improving forecasting accuracy and eliminating possible cyber-attacks. \\
As future work, we are planning to build deep-learning based mitigation methods for the adversarial machine learning attacks such as FGSM, BIM, and DeepFool. In order to identify malicious inputs, there are some anomaly detection methods, for example, AutoEncoders or One-Class support vector machine. The exploration of the feasibility of these anomaly detection and outlier methods can also be considered as one of the potential future work domains which might be based on the findings of this study. 
The proposed concept has potential to inspire other researchers and engineers working on edge analytics, cybersecurity, especially in the field of energy, data science, and similar digitalization domains who aim to create next-generation implementations and methods in this emerging field.}

\bibliographystyle{plain}
\bibliography{references.bib}

\end{document}